\documentclass{PoS}

\usepackage{comment}
\usepackage{amsmath}
\usepackage{amsfonts}
\usepackage{amssymb}%

\newcommand{\be}{\begin{equation}}
\newcommand{\ee}{\end{equation}}
\newcommand{\bdm}{\begin{displaymath}}
\newcommand{\edm}{\end{displaymath}}
\newcommand{\btb}{\begin{tabular}}
\newcommand{\etb}{\end{tabular}}
\newcommand{\bfig}{\begin{figure}}
\newcommand{\efig}{\end{figure}}

\newcommand{\I}{\mathrm{Im}}

\title{Radially excited $\psi$ mesons and the $Y$ enhancements}

\ShortTitle{$\psi$ mesons and the $Y$ enhancements}

\author{\speaker{Susana Coito}\\
        Institute of Physics, Jan Kochanowski University, 25-406 Kielce, Poland\\
        E-mail: \email{scoito@ujk.edu.pl}}

\abstract{While many properties of the vector charmonium first excitations are yet to be measured, enhancements at unexpected energies are intriguing, alias the $Y$ states. In order to understand the naturally unquenched mesonic line-shapes, the influence of the most relevant hadronic decay channels must be taken into account. Within an unitary effective approach we present results where mesonic loops are included in an equivalent manner to coupled-channels. We show results for the $\psi(3770)$ and $\psi(4160)$ systems, where we find the nonperturbative effects of dynamical generation of poles and line-shape distortion. }

\FullConference{XIII Quark Confinement and the Hadron Spectrum - Confinement2018\\
		31 July - 6 August 2018\\
		Maynooth University, Ireland}

\begin{document}

%%%%%%%%%%%%%%%%%%%%%%%%%%%%%%%%%%%%%%%%%%%%%%%%%%%%%%%%%

\section{Introduction}

The charmonium vector states, alias the $\psi$ states, can only be considered strictly ``quenched'', i.e., almost pure $\bar{c}c$ states, if they lay below all the hadronic decay channels involving open-charm mesons, which are allowed by conservation laws. In this category fall only the $J/\psi$ and the $\psi(2S)$ mesons \cite{pdg}. All the higher $\psi$'s are strongly influenced by their nearby decay channels, not only the open ones, but the closed ones as well. Moreover, through the same decay channels, they are also strongly influenced by the nearest other $\psi$'s, below and above. A comprehensive study of the charmonium vector spectrum should then include all the interferences that are expected from the simplest interactions, which is an extensive and intricate quest. Some of the phenomena that can result from such interferences are i) deformation of the line-shapes, ii) generation of poles from the continuum, iii) shifting of the mass peak positions in certain decay channels due to special interferences between the background and a certain opening threshold. 

\begin{figure}
\begin{center}
\resizebox{!}{340pt}{\includegraphics{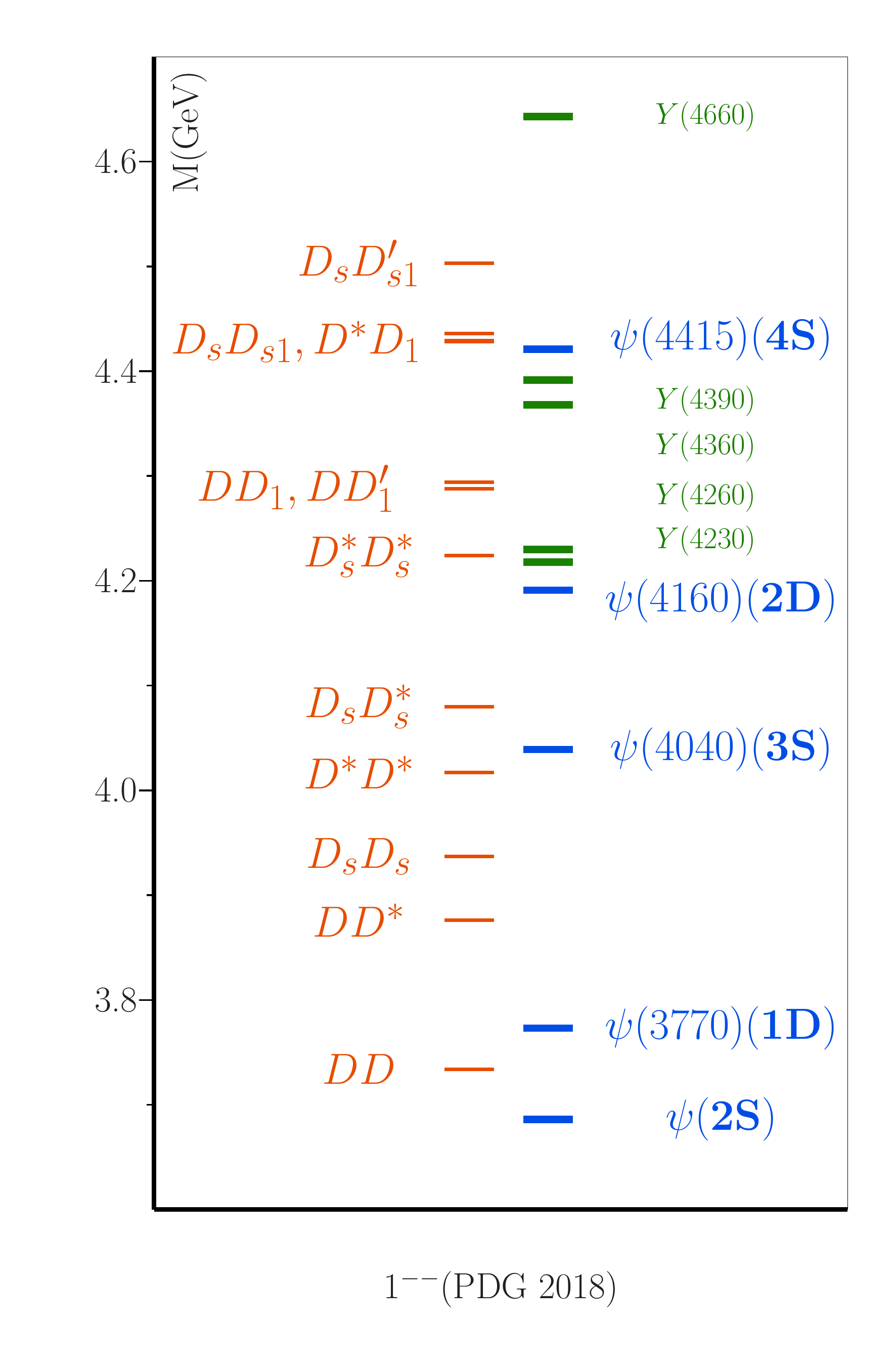}}
\caption{\label{fig1} Vector charmonium spectrum, position of the $Y$ enhancements, and dominant decay modes.}
\end{center}
\end{figure}

The understanding of some of these interferences should shed light over an apparently different problem, that is the emergence of $Y$ enhancements, i.e., vectorial structures appearing in the charmonium energy region in certain non-dominant hadronic decay modes, such as the $J/\psi\pi^+\pi^-$ \cite{prl118p092001,prl110p252002}, $h_c\pi^+\pi^-$ \cite{prl118p092002}, or $\psi(2S)\pi^+\pi^-$ \cite{prd91p112007}, with different masses than the $\psi$ states, cf.~Fig.~\ref{fig1}. While some of the $Y$ signals can actually be mostly independent from the $\psi$ resonances, others can be due to interferences generated by them, and such possibilities are worthy to be analyzed.      

We pay attention to the fact that a resonance is not determined by a Breit-Wigner fit to a peak in the mass distribution to a certain decay channel. For instance, a rough look to data in channels $DD$, $DD^ *$, and $D^*D^*$, both from BaBar \cite{prd79p092001} and Belle \cite{prd77p011103}, allows us to see that the peaks and dips are in different positions in the different channels and therefore, it would be simply wrong to fit with Breit-Wigners to each peak in each decay channel. Instead, what has been done to determine the mass of the $\psi$ states, was the analysis of the $R$ distribution, i.e., the ratio of the cross section to all hadrons and the cross section to leptons \cite{plb660p315,prd72p017501}. An equivalent analysis using non-dominant hadronic decay amplitudes only, would be helpful to a better identification of the $Y$ states. 

In this work we show results of an unitarized effective Lagrangian model that has been previously applied to the light meson sector to generate the dynamical resonances $\kappa$ and $a_0(980)$ \cite{prd93p014002,npb909p418}. We hypothesize that a similar phenomena could occur to higher energy systems. Indeed, different works have contemplated such possibilities, e.g.~in \cite{prl91p012003} for an open-charm scalar $D_0^*(2100-2300)$, or in \cite{prd76p074016,epja36p189}, for a charmonium scalar at 3.7 GeV. A material indication that the latter resonance exists may be found in Ref.~\cite{prl100p202001}, where Belle has observed different peaks in the different channels $DD$, $DD^*$, and $D^*D^*$ compatible with scalar quantum numbers. Rather than different resonances, one can consider only one pole below $DD$ threshold that generates a tail in the line-shape of each decay channel, in which case the assignment of a different state $X$ to each one of the peaks in the different invariant masses is, as discussed above, incorrect. 

In Refs.~\cite{appbs10p1049,1712.00969} we studied in detail the line-shape of the $\psi(3770)$ within the model we briefly discuss here, where in addition to a seed pole, a dynamically generated pole has been found. Here, we present some of our main results related to the $\psi(3770)$. Concerning the higher states, we show line-shape results for the $\psi(4160)$, following the discussion started in Refs.~\cite{posh2017p030,1810.03532}. Moreover, we refer to the work in Refs.~\cite{piotrowska,1810.03495}, where the $\psi(4040)$ has been treated with the same approach, in which the authors have found a dynamical pole around 4.0 GeV.

\section{\label{model}The model}

We employ an unitarized effective Lagragian model to compute the line-shape and cross section of production processes of the type $e^+e^-\to\psi\to m_1m_2$, where $m_{1,2}$ is the final meson$-$anti-meson system. Rather than a simple massive vector in the intermediate state, we consider meson-meson channels in the form of one-loop that obey the ``Born'' expansion, i.e., the occurrence probability of one one-loop is higher than of two one-loops, etc.. Thus the scalar part of the propagator series is convergent and obeys a geometric progression that sums up to     
\be
\label{prop}
\Delta_\psi(s)=\frac{1}{s-m_{\psi}^{2}+\sum_j^N\Pi_j(s)},
\ee
where $m_\psi$ is the bare mass of the $\psi$, $s$ is the invariant energy squared, $N$ is the number of channels, and $\Pi_j(s)$ is the loop function of channel $j$, that is given by
\be
\Pi_j(s)=\Omega_j(s) + i\sqrt{s}\Gamma_j(s),\ \ \Omega,\ \Gamma \in \Re\ .
\ee
The real part of $\Pi_j$ is obtained from the dispersion relations
\be
\Omega_j(s,m_1,m_2)=\frac{PP}{\pi}\int_{s_{thj}}^{\infty}\frac{\sqrt{s'}\Gamma_j(s^{\prime},m_1,m_2)}%
{s^{\prime}-s}\ \mathrm{d}s^{\prime},
\ee
and the imaginary part is given by 
\be
\Gamma_{\psi\to(m_1m_2)_j}(s)=\frac{k_j(s,m_1,m_2)}{8\pi s}
|\mathcal{M}_{\psi\to(m_1m_2)_j}|^{2}\ ,
\ee
where $k_j$ is the final state momentum, and
\be
\label{amp}
|\mathcal{M}_{\psi\to(m_1m_2)_j}|^{2}=\mathcal{V}_j(s,m_1,m_2)f_{\Lambda}^{2}(q_j^2).
\ee
The function $\mathcal{V}$ is the vertex amplitude, computed using the Feynmann rules, and $f$ is a vertex form factor that depends on a cutoff parameter $\Lambda$, and on the off-shell momentum $q_j$, and it is here defined by an exponential as
\be
\label{ff}
f_{\Lambda}(\vec{q}_j^2)=e^{-q_j^2/\Lambda^2}.
\ee
We note that our full vertex function is the product of $\mathcal{V}$ with $f^2$, and therefore the ``form-factor'' $f$ is conceptually different than the traditional electromagnetic form-factor that defines the full vertex in a model independent way. The spectral function is given in function of the energy $E=\sqrt{s}$ by 
\be
\label{sf}
d_{\psi}(E) =-\frac{2E}{\pi}\mathrm{Im}\ \Delta_\psi(E),
\ee
which due to unitarity comes automatically normalized to 1.

\section{Results}

\subsection{The $\psi(3770)$}

\begin{figure}
\begin{center}
\resizebox{!}{200pt}{\includegraphics{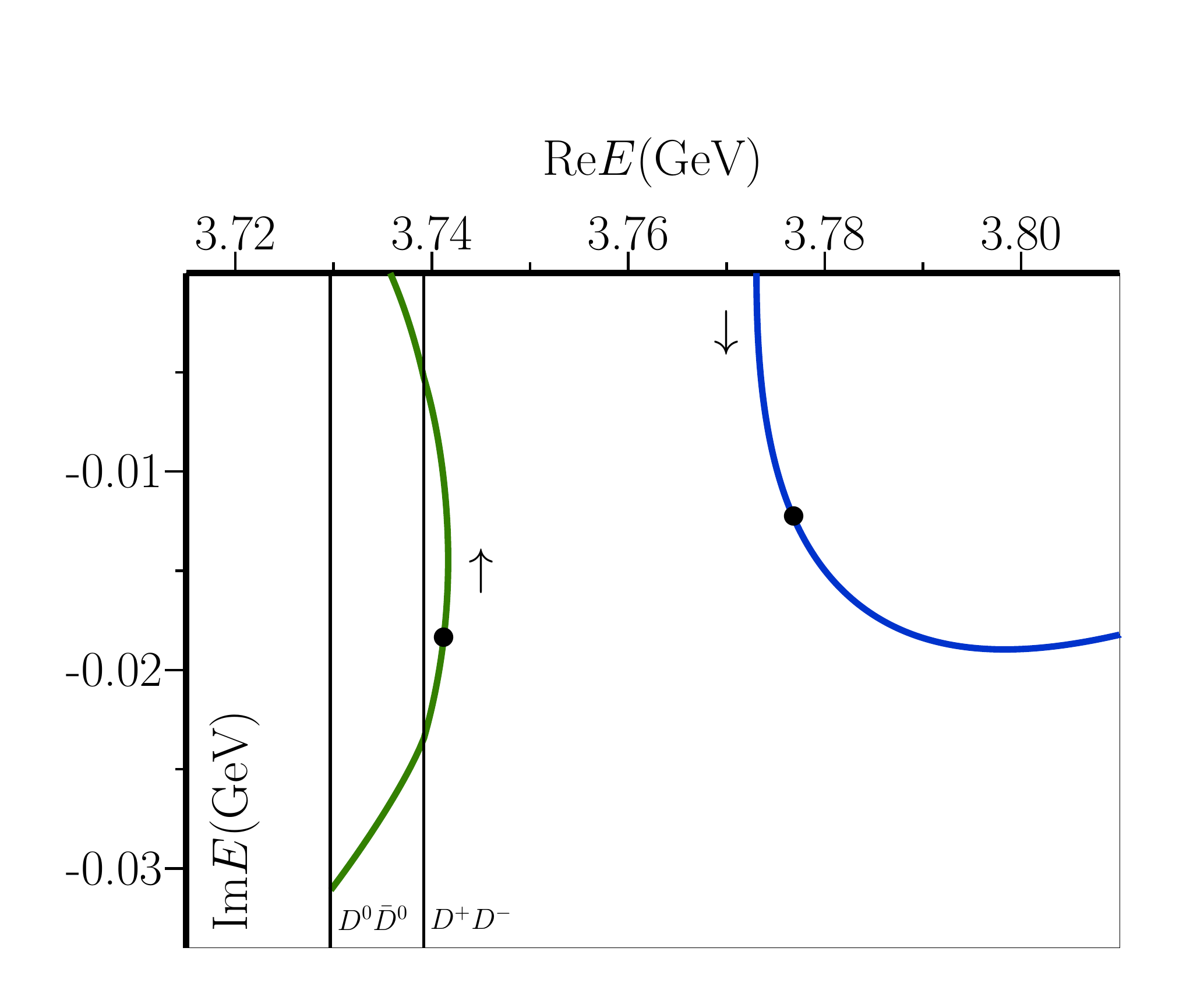}}
\caption{\label{fig2} Pole trajectories of the $\psi(3770)$. Left green line: trajectory of the dynamical pole, right blue line: trajectory of the seed pole. The arrows indicate the direction of increasing the coupling at vertices $\psi DD$. Bullets: pole positions for the fitting parameters (cf.~text and Ref.~ \cite{1712.00969}).}
\end{center}
\end{figure}

A detailed study of the $\psi(3770)$ can be found in Refs.~\cite{1712.00969,appbs10p1049}. Here we present our main results. The interaction Lagrangian density is defined by  
\be
\mathcal{L}_{\psi (P_1P_2)_j}=ig_{\psi (P_1P_2)_j}\psi_{\mu}\sum_j^2\Big((\partial^{\mu}P_1P_2)_j-(\partial^{\mu}P_2P_1)_j\Big),
\ee
where $P_1P_2$ are the pseudoscalar mesons $D^+D^-$ and $D^0\bar{D}^0$. The amplitude $\mathcal{V}$ comes as
\be
\mathcal{V}_j(s,P_1,P_2)=g_{\psi (P_1P_2)_j}^{2}\frac{4}{3}k_j^{2}(s,P_1,P_2).
\ee
Poles are found when the denominator of the propagator (Eq.~\eqref{prop}) is zero on the proper Riemann sheet, defined by the condition $\I\ k_j<0$ for $E>\sqrt{s_{th_j}}$ in both channels. Only three free parameters are needed, the cutoff $\Lambda$, the seed mass $m_\psi$, and the effective coupling for the vertices $g_{\psi (P_1P_2)}\equiv g_{\psi DD}$, which is the same for both channels. The three parameters are obtained by the fit to cross section data in Ref.~\cite{plb668p263} with $\chi^2\simeq 1.03$ and we get $\Lambda \simeq 272$ MeV, $m_\psi\simeq 3773$ MeV, and $g_{\psi DD}\simeq 31$. Two poles were found, at $3741-i18$ MeV and at $3777-i12$ MeV, the first one from the continuum, and the second one coming from the seed, as it can be see in Fig.~\ref{fig2}. The existence of these two poles in the same Riemann Sheet is not evident from the line-shape that shows one peak only, but distorted with relation to a Breit-Wigner shape (cf.~Fig.~3 of Ref.~\cite{posh2017p030}).  

\subsection{The $\psi(4160)$}

\begin{figure}
\begin{center}
\resizebox{!}{200pt}{\includegraphics{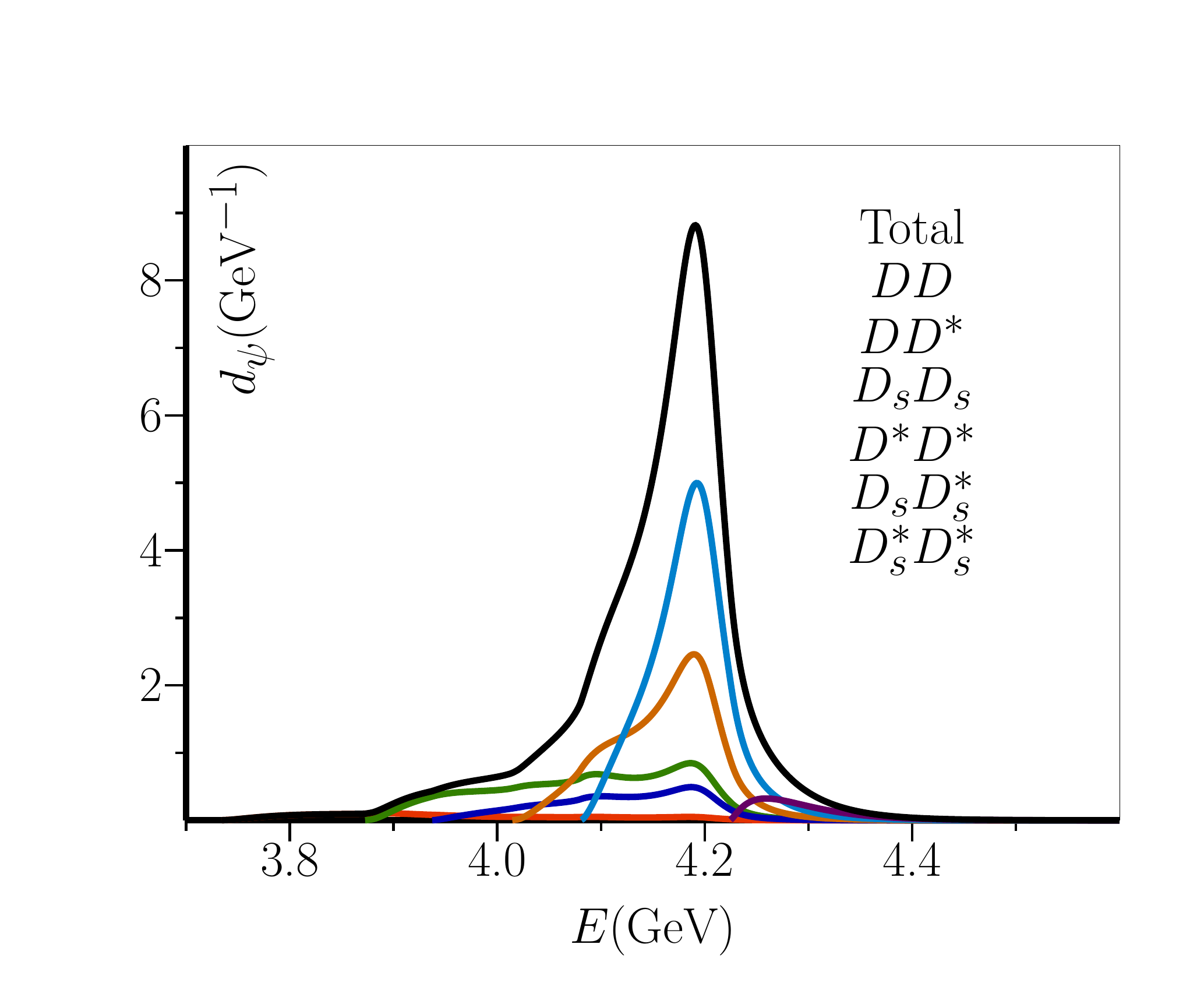}}
\caption{\label{fig3} Normalized spectral function of the $\psi(4160)$ (see Eq.~\eqref{sf}) for $\Lambda=450$ MeV (black), and partial spectral functions. In threshold opening order, $DD$ (red), $DD^*$ (green), $D_sD_s$ (dark blue), $D^*D^*$ (orange), $D_sD_s^*$ (light blue), and $D_s^*D_s^*$ (violet).}
\end{center}
\end{figure}

In a similar way, we employ the model to the vector $\psi(4160)$ that, as the $\psi(3770)$, should be dominantly a $d$-wave. The system is more complex than the $\psi(3770)$ though, since there are more possibilities of decay (see Fig.~\ref{fig1}). For simplicity, we include only the open channels with open charm, viz.~$DD$, $DD^*$, $D^*D^*$, $D_sD_s$, $D_sD_s^*$, and $D_s^*D_s^*$. We consider the coupling strengths to be flavor independent, and therefore we use the partial decay ratios $\Gamma_{DD^*}/\Gamma_{D^*D^*}\sim 0.34$ and $\Gamma_{DD}/\Gamma_{D^*D^*}\sim 0.02$ in \cite{pdg} to compute all the partial couplings involved at vertices $\psi (m_1m_2)_j$. The relative amplitudes to each channel are shown in Fig.~\ref{fig3}, together with the total amplitude through the line-shapes, for a cutoff $\Lambda=450$ MeV. The remaining free parameter is the seed mass which is set to reproduce the peak mass and width at the values given by PDG \cite{pdg}, i.e. $m\sim 4191$ MeV and $\Gamma\sim 70$ MeV. For this cutoff it comes $m_\psi\simeq 4154$ MeV, a value that is thus shifted upwards by the contribution of all the included one-loops, following a similar behavior as the seed pole in Fig.~\ref{fig2}. From Fig.~\ref{fig3}, the dominant decay channel is the $D_sD_s^*$, but its branching ratio has not yet been determined in the experiment. We can also verify that the peak position is the same in every channels, although the line-shape of the partial amplitudes change. In Ref.~\cite{1810.03532} we discussed the line-shape of the decay $\psi(4160)\to J/\psi f_0(980)$ with the one-loop effect only, using the six channels above, concluding that the peak position would not shift, therefore no peak for the $Y(4260)$ could be obtained. The fact that in the experiment a peak is not seen at the $\psi(4160)$ position, in the mode $J/\psi\pi^+\pi^-$ \cite{prl118p092001,prl110p252002}, could be justified by a very small coupling to this channel. Yet, given the large phase space available, it is unexpected that the $\psi(4160)$ does not manifest in this mode. Instead, one may consider the existence of some other effect that would actually shift the $\psi(4160)$ position, specifically in the $J/\psi f_0(980)\to J/\psi \pi^+\pi^-$ channel, leading even to the line-shape of the $Y(4260)$. Such idea is developed in Ref.~\cite{prep}.  

\begin{figure}
\begin{center}
\resizebox{!}{200pt}{\includegraphics{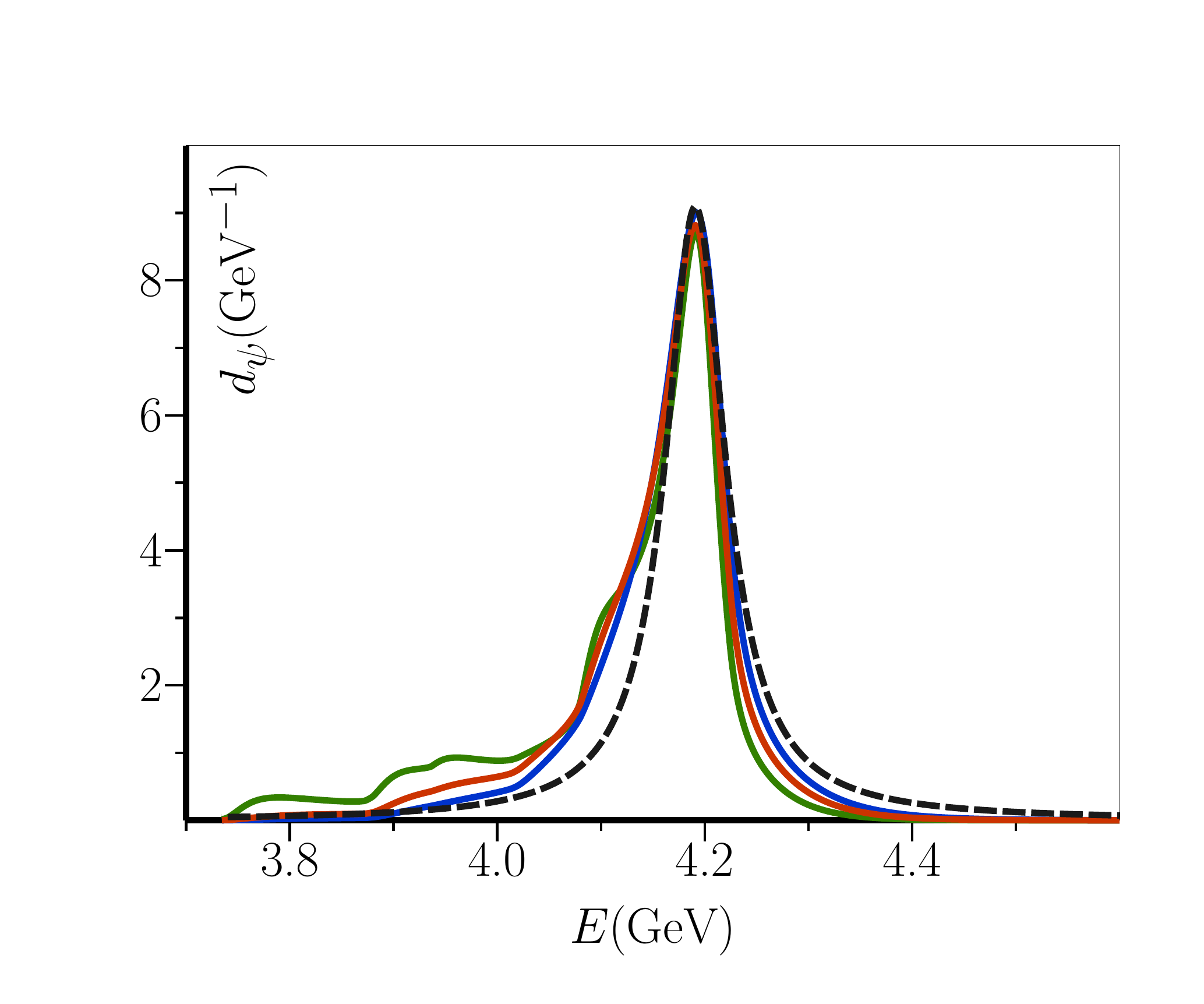}}
\caption{\label{fig4} Variation of the normalized spectral function of the $\psi(4160)$, with the cutoff $\Lambda$. Green line: $\Lambda=400$ MeV, red line: $\Lambda=450$ MeV, blue line: $\Lambda=500$ MeV, black dashed line: Breit-Wigner distribution.}
\end{center}
\end{figure}

Finally, in Fig.~\ref{fig4} we study the variation of the total line-shape with the cutoff. For smaller cutoff parameters, the distortion with relation to a Breit-Wigner line-shape is more pronounced. This can be understood from Eq.~\eqref{ff}, from where the exponential falls off quicker for smaller $\Lambda$ values, thus limiting the range of influence of each loop to smaller energies. Away from the resonance peak, one should include other components such as the $\psi(4040)$ and the $\psi(3770)$ on the left, and the $\psi(4415)$ on the right. Our result is thus qualitative, so as the value of the cutoff $\Lambda$, which differs from the value obtained for the ``cleaner'' $\psi(3770)$ system, that came from the fit. 
%A study of the poles in the vicinity of the $\psi(4160)$ may be found in \cite{prep}.

\section{Conclusion}

In conclusion, we aim to disentangle whether some of the $Y$ enhancements can be manifestations of the $\psi$ states. We presented results concerning the $\psi(3770)$ and $\psi(4160)$ vectors, within an unitarized effective Lagrangian model, where the most relevant one-loop channels have been included. A distortion from a Breit-Wigner line-shape is visible for both $\psi$'s. A dynamically generated companion pole is found for the $\psi(3770)$, yet not leading to an independent peak in the line-shape. The poles of the $\psi(4160)$ have not been examined here. The cutoff parameter is regarded as a free parameter, to which one can attribute a physical meaning, related to the size of the system, with the possibility that it is also an effective parameter, i.e.~that it somehow depends on the number of variables included in the problem. The one-loop dynamics only leads to an overall shifting from the seed mass, but not to the displacement of the peak position in different channels. Therefore, the $Y(4260)$ does not rise as a manifestation of the $\psi(4160)$ within this model version. Other mechanisms to address the origin of the $Y(4260)$ are under study.  

%As for the shifting of the peak position in different channels, other interferences should be studied, since the one-loop dynamics only leads to an overall shifting from the seed mass, but not to the displacement of the peak position in different modes.\\

\section*{Acknowledgements}

The author thanks to F.~Giacosa and to K.U.~Can for useful discussions. This work was supported by the
\textit{Polish National Science Center} through the project OPUS no.~2015/17/B/ST2/01625.

\end{document}